\documentclass[12pt]{article}

\usepackage{color}
\usepackage{scicite}
\usepackage{times}

\usepackage{graphicx}
\usepackage{soul}

\newenvironment{sciabstract}{%
\begin{quote} \bf}
{\end{quote}}

\title{Origami-based impact mitigation via rarefaction solitary wave creation}

\author
{Hiromi Yasuda,$^{1}$ Yasuhiro Miyazawa,$^{1, 2}$ Efstathios G. Charalampidis,$^{3}$ \\ Christopher Chong,$^{4}$ Panayotis G. Kevrekidis,$^{3}$ Jinkyu Yang$^{1\ast}$\\
\\
\normalsize{$^{1}$Department of Aeronautics \& Astronautics, University of Washington,}\\
\normalsize{Seattle, WA 98195-2400, USA}\\
\normalsize{$^{2}$Department of Mechanical and Aerospace Engineering, Tohoku University,}\\
\normalsize{Sendai, Miyagi 980-8577, Japan}\\
\normalsize{$^{3}$Department of Mathematics and Statistics, University of Massachusetts,}\\
\normalsize{Amherst, MA 01003-4515, USA}\\
\normalsize{$^{4}$Department of Mathematics, Bowdoin College,}\\
\normalsize{Brunswick, ME 04011, USA}\\
\\
\normalsize{$^\ast$To whom correspondence should be addressed; E-mail:  jkyang@aa.washington.edu.}
}

\date{}

\begin{document} 


\baselineskip24pt


\maketitle


\begin{sciabstract}
  The principles underlying the art of origami paper folding can be applied to design sophisticated metamaterials with unique mechanical properties.
 By exploiting the flat crease patterns that determine the dynamic folding and unfolding motion of origami, we are able to design an origami-based
metamaterial that can form rarefaction solitary waves. Our analytical, numerical and experimental results demonstrate that this rarefaction solitary wave overtakes initial compressive strain waves, thereby causing the latter part of the structure to feel tension first instead of compression. This counter-intuitive dynamic mechanism can be used to create a highly efficient--yet reusable--impact mitigating system without relying on material damping, plasticity or fracture.
 \end{sciabstract}

 Mechanical metamaterials offer a new dimension in achieving nonconventional and tailored mechanical properties through architecture~\cite{Lee2012,Wegener2013,Paulose2015,Ma2016,Coulais2017}.
 In particular, the manipulation of wave propagation in mechanical metamaterials is a topic of intense research for various engineering applications, e.g., waveguiding, vibration filtering, subwavelength imaging, and impact mitigation \cite{Zhang2009,Shan2015,Babaee2016,Lee2016,Chen2016,Bilal2017}. 
 The ability to achieve such desirable mechanical performance often relies on the platform in which we construct mechanical metamaterials.
 The choice of platform can vary from periodically arranged micro-/macro-lattice structures, self-assembling particles, to 3D printed soft/hard architected materials~\cite{Nesterenko2001,Hiraiwa2016,Raney2016, Deng2017,vakakis2017}.

 Recent studies have shown that origami can serve as an ideal playground to realize highly versatile and tunable mechanical metamaterials~\cite{Schenk2013,Silverberg2014,Lv2014,Felton2014,Waitukaitis2015,Silverberg2015,Overvelde2016}.
 {For example, by introducing crease lines into flat surface materials, one can construct origami-based structures which can offer enhanced stiffness\cite{Filipov2015}, negative Poisson's ratio\cite{Yasuda2015}, and multi-stability \cite{Hanna2014,Jianguo2015,Yasuda2017}.}
 Given the scale-free nature of origami, this kind of design framework can be utilized in a wide range of scales. 
 For example, the concept of origami has been adapted to a diverse set of design principles, including robotics~\cite{Miyashita2015}, reconfigurable structures~\cite{Overvelde2017}, and self-folding actuated by living-cells~\cite{Kuribayashi-Shigetomi2012}.
While these studies focused mainly on origami's static or quasi-static behavior, the analysis of the dynamics of origami-based structures is a natural next step to investigate. However, the connection between the origami crease pattern and the dynamic folding/unfolding behavior of origami itself has been relatively unexplored~\cite{Ma2014,SchenkDynamic2014}.
 In particular, very few experimental studies have been reported~\cite{Tsuda,Zhou2016}.
 
\begin{figure} 
\centering
\includegraphics[width=0.9\linewidth]{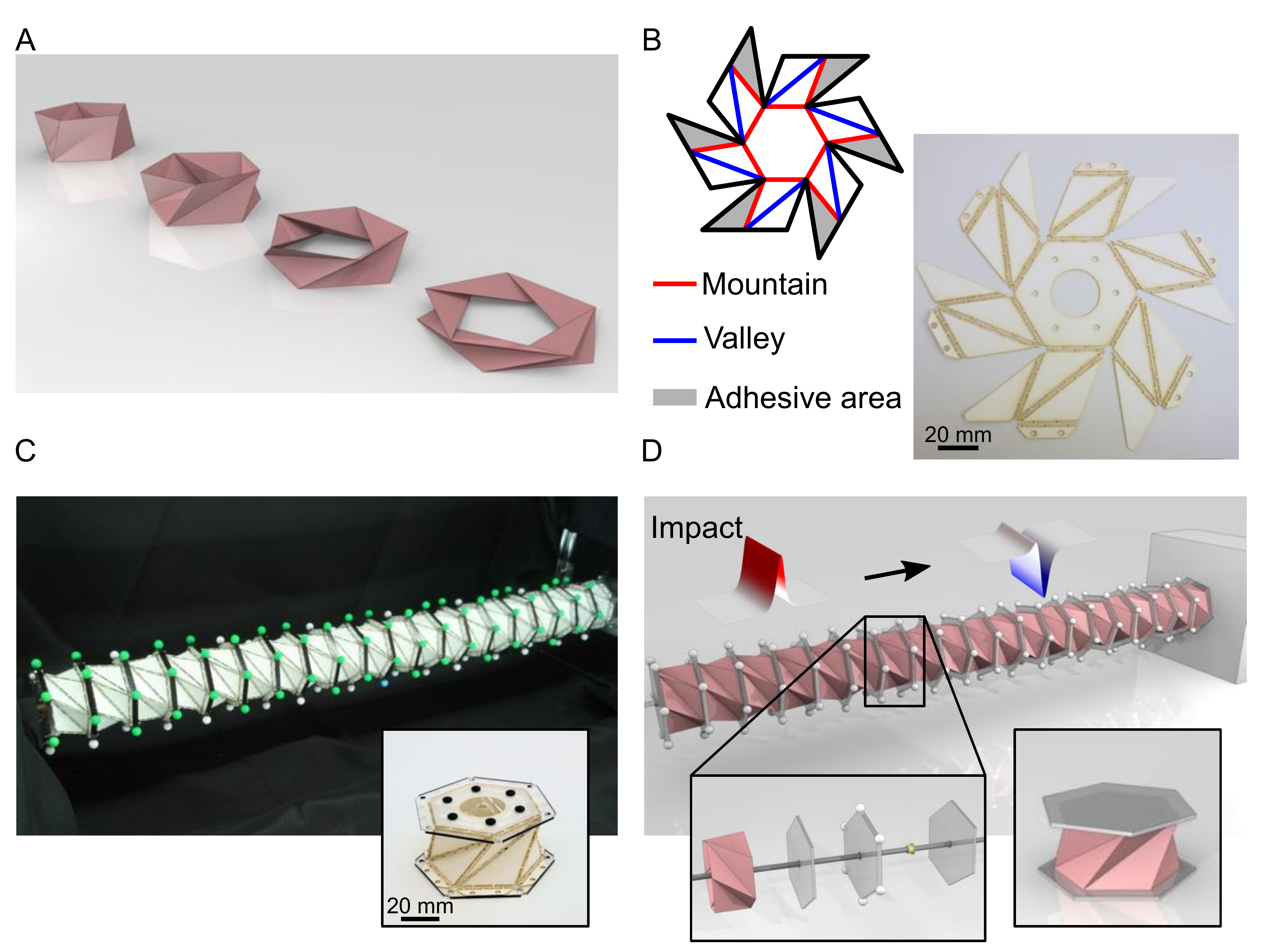}
\caption{ \textbf{Geometry of the TCO prototypes.} (\textbf{A}) Folding motion of the TCO is shown in sequence. (\textbf{B}) The flat sheet with crease patterns (upper left) is composed of mountain crease lines (red), valley crease lines (blue), and the adhesive area (shaded area). The photograph shows corresponding laser-cut paper sheets (lower right). (\textbf{C}) The actual prototype of the origami-based metamaterial and its unit cell (lower right inset). (\textbf{D}) The origami-based metamaterial generates the rarefaction solitary wave despite the application of compressive impact. The system is composed of the TCO unit cells (lower right). To connect the neighboring unit cells, the interfacial polygonal cross-section with markers at vertices is used (lower left). }
\label{fig:Concept}
\end{figure}

 In the present study, we explore unique wave dynamics  in a mechanical metamaterial that is composed of volumetric origami structures.
 In particular, each volumetric origami structure is a Triangulated Cylindrical Origami (TCO) (Fig.~\ref{fig:Concept}A for the folding motion and B for the flat crease pattern before it is assembled into a unit cell)~\cite{Yasuda2017}.
 This TCO unit cell is analogous to a post-buckled shape of a cylinder under simultaneous axial and twisting loading~\cite{Guest1994,Hunt2005}. Thus, the folding motion of the TCO is characterized by coupling between axial and rotational motions as shown in Fig.~\ref{fig:Concept}A.
 Recent studies on the TCO structure have shown their versatile mechanical properties, such as its tailorable stability \cite{Jianguo2015}, zero-stiffness mode \cite{Ishida2014}, and strain-softening/hardening behavior \cite{Yasuda2017}.

Our origami structure consists of twenty TCO unit cells, which are fabricated by using paper sheets cut by a laser cutting machine (Figs.~\ref{fig:Concept}, B and C, see also Materials and Methods for details). The crease patterns are carefully designed to make the TCO cells exhibit effective strain-softening behavior, as in~\cite{Yasuda2017}.
 This softening behavior in origami is in sharp contrast to conventional nonlinear metamaterials that mainly exploit strain-hardening behavior, e.g., granular crystals that have been serving as a popular testbed for demonstrating classical Fermi-Pasta-Ulam-Tsingou dynamics~\cite{Nesterenko2001,vakakis2017}. 
 

 Recently, it has been theoretically and numerically predicted that a one-dimensional (1D) discrete system with strain-softening interactions can support the propagation of a solitary wave, in the form of a rarefaction pulse~\cite{Nesterenko2001,Herbold2013,Yasuda2016}.
 In this 1D system, the application of a compressive impact generates a tensile solitary wave, propagating ahead of the initial compressive strain wave and thereby making the latter part of the medium feel tension first instead of compression (see Fig.~\ref{fig:Concept}D for the conceptual illustration). While the feasibility of verifying such  counter-intuitive dynamics has been discussed in different settings~\cite{Fraternali2014,Yasuda2016}, the experimental demonstration of such rarefaction solitary waves  has remained elusive. 
Here, we report the first experimental observation of a mechanical rarefaction solitary wave
that overtakes the leading compressive wave generated by a compressive impact. The experimentally observed rarefaction pulse agrees quantitatively with numerical simulations of the derived model and agrees qualitatively with an asymptotic
analysis based on the celebrated Korteweg--de Vries (KdV) equation.


\section*{Static analysis}
\begin{figure} 
\centering
\includegraphics[width=0.7\linewidth]{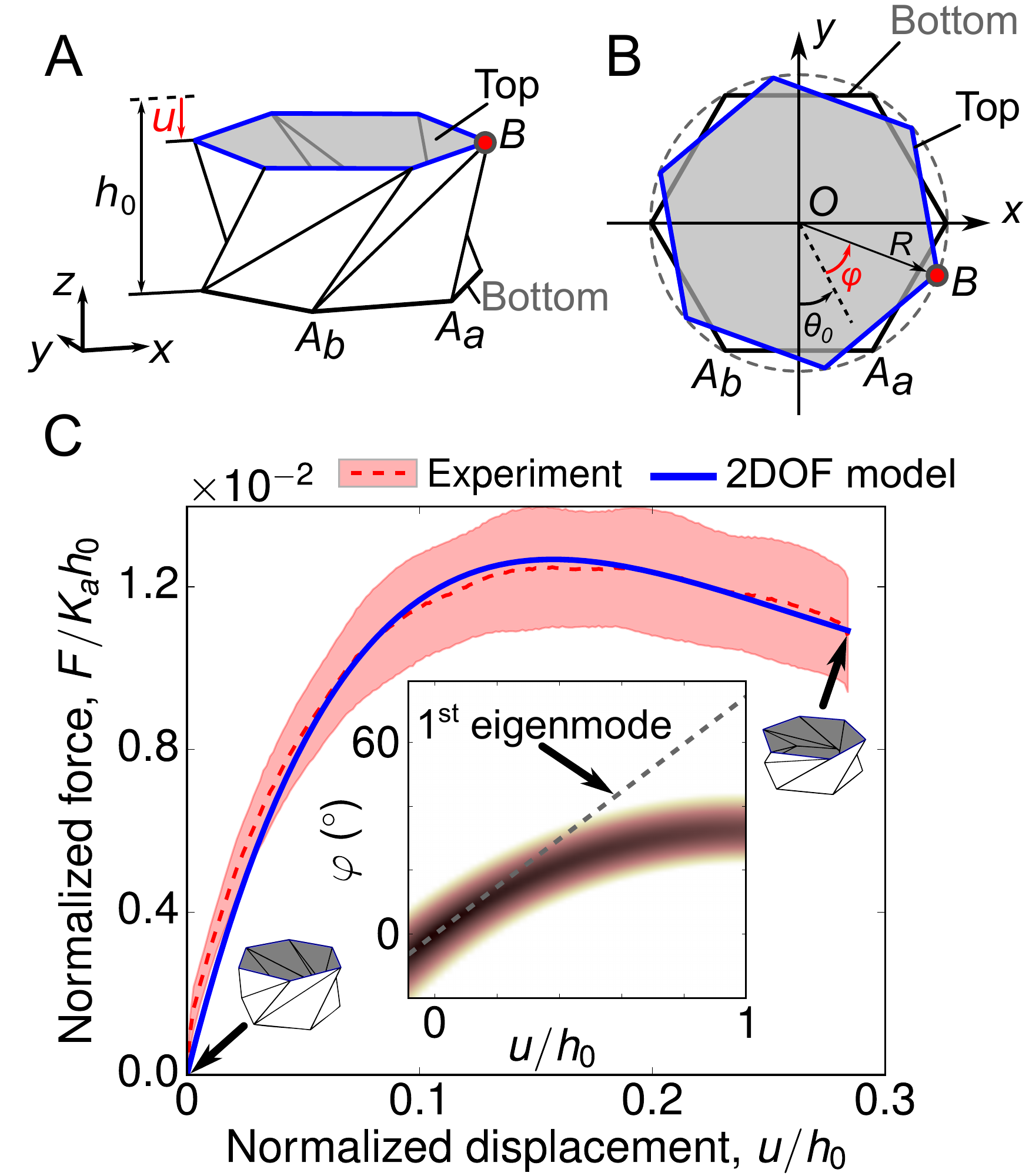}
\caption{ \textbf{Folding motions of the TCO with strain softening behavior.} (\textbf{A}) The axial displacement ($u$) is defined with respect to the initial height ($h_0$) of the TCO. (\textbf{B}) Top-down view shows the rotational angle ($\varphi$) defined with respect to the initial angle ($\theta_0$). (\textbf{C}) Experimentally measured axial force ($F$) normalized by the spring constant ($K_a$) and $h_0$ (the mean value is shown as a dashed curve, and the standard deviation is represented by colored areas) is compared with the 2DOF linear spring model (blue). The inset shows the surface plot of the elastic potential energy. The darker color indicates the lower energy level. The dashed line shows the folding behavior of the reduced 1D model obtained from the first eigenmode.}
\label{fig:Static}
\end{figure}

 We start by examining the geometry of the TCO unit cell.
 The TCO structure has axial and rotational motions that are coupled to each other. We characterize this folding motion by using the axial displacement ($u$) and rotational angle ($\varphi$), which are defined with respect to the initial height ($h_0$) and angle ($\theta_0$) (Figs.~\ref{fig:Static}, A and B).
 To analyze the kinematics of the origami, we construct a two degree-of-freedom (DOF) mathematical model which approximates the folding behavior of the TCO into linear spring motions along the crease lines \cite{Yasuda2017}.
 In this model, we use two different spring constants ($K_a$ and $K_b$) for the shorter ($\overline{A_aB}$ in Fig.~\ref{fig:Static}A) and longer crease lines ($\overline{A_bB}$), respectively.
 These two different spring constants ($K_a$ and $K_b$) are repeated on a single unit cell.
 The creases (e.g., $\overline{A_aA_b}$) along the edge of the $N_p$-sided polygon
  (denoted by the shaded area in Figs.~\ref{fig:Static} A and B with the radius $R$ of the circle circumscribing the polygon) are modeled as a torsional spring with the spring constant $K_{\psi}$.
 These three spring constants are determined empirically by conducting compression tests on the fabricated TCO cells (see Materials and Methods for details. Also see Supplementary Movie S1 for the fabrication and folding motion of the TCO).
  
  We analyze the elastic potential energy change as a function of $u$ and $\varphi$ as follows:
\begin{equation} \label{eq:ElasticPotential}
{U}\left( {u}, \varphi \right)=\frac{1}{2}{{N}_{p}}{{K}_{a}}{{\left( {a}-a^{(0)} \right)}^{2}}+\frac{1}{2}{{N}_{p}}{{K}_{b}}{{\left( {b}-b^{(0)} \right)}^{2}}+\frac{1}{2}(2{{N}_{p}}){{K}_{\psi }}{{\left( {{\psi }}-\psi^{(0)} \right)}^{2}}
\end{equation}
where $a$ and $b$ are the length of the crease lines ($\overline{A_aB}$ and $\overline{A_bB}$) respectively, and the superscript $(0)$ denotes the initial states.
 By using this energy expression and applying  the minimum potential energy principle~\cite{Yasuda2017}, we obtain the near-curve trajectory which indicates how the deformation of the TCO cell takes place by coupling axial and rotational motions (see the inset of Fig.~\ref{fig:Static}C, where the dark color indicates a minimal energy regime). 

 
In this study, we prototype twenty identical TCO unit cells by choosing $h_0$ = 35 mm, $\theta_0$ = $70^\circ$, $R$ = 36 mm, and $N_p$ = 6.
 To ensure uniform and repeatable folding behavior in these units, we apply the preconditioning process to each cell and verify the uniformity by measuring its compressive motions (see Section I.A in Supplementary Materials for details).
 Figure \ref{fig:Static}C shows the experimental result (solid curve), along with the analytical prediction (dashed curve) obtained from the 2DOF model in Eq. (1).
 We observe the analytical model corroborates the experimental result.
 Figure \ref{fig:Static}C also confirms the strain softening behavior of the TCO under compression, i.e., the stiffness of the structure decreases as it is compressed.
 
\section*{Dynamic analysis}
\begin{figure} 
\centering
\includegraphics[width=1.\linewidth]{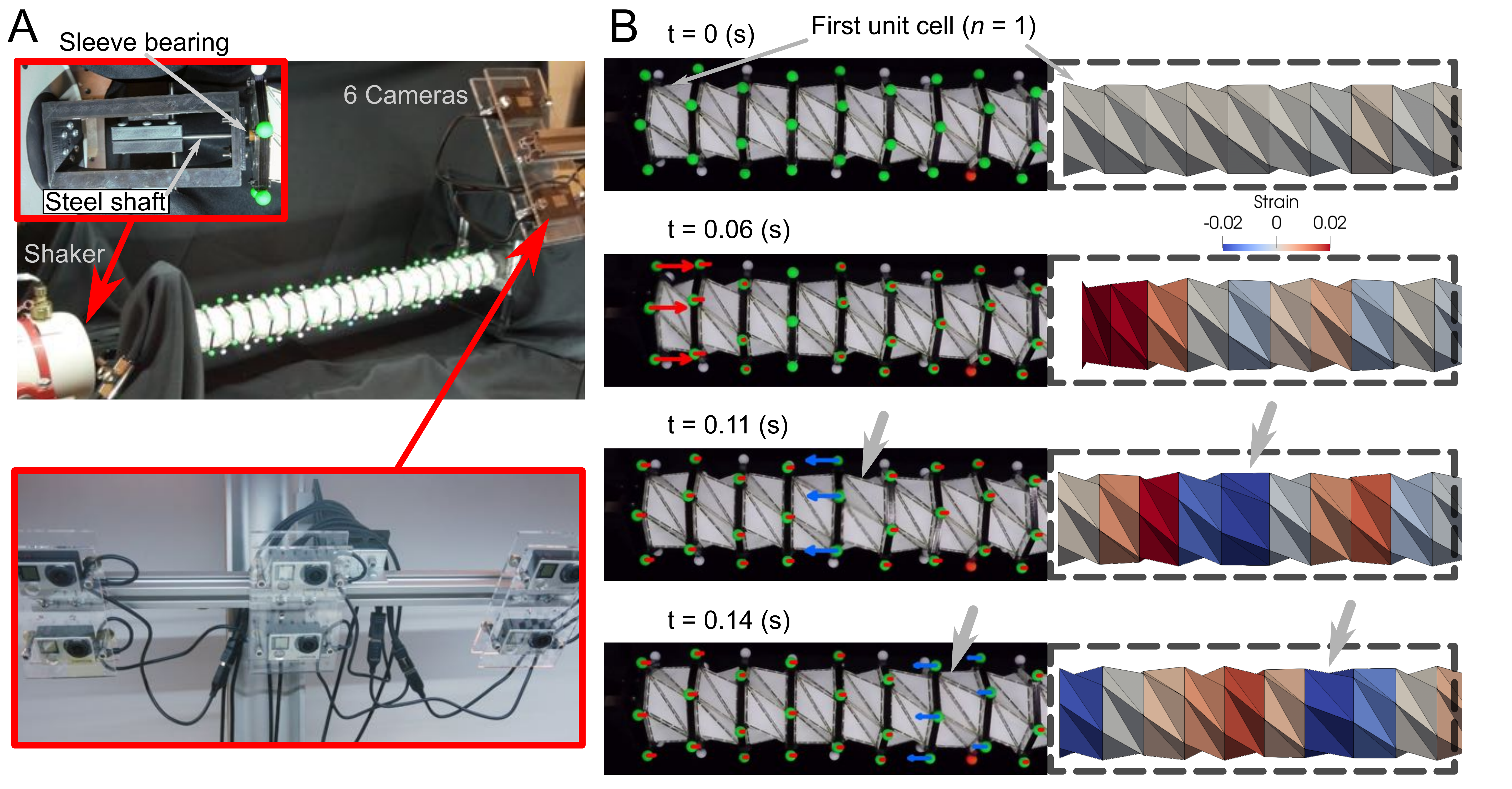}
\caption{ \textbf{Experimental setup and DIC analysis results.} (\textbf{A}) The shaker is attached to the left-most unit cell through the sleeve bearing (upper left inset). The folding motion of each unit cell is captured by six action cameras (lower inset). For DIC analysis, the fluorescent green markers are used. (\textbf{B}) Snapshots of the experiment at $t$ = 0, 0.06, 0.11, 0.14 s. Images from the camera are shown in the left column where the arrow represents the velocity vector of the polygon in the axial direction. 3D reconstruction of the TCO chain (right column). The deformation is scaled 2.5 times larger than the original deformation for visual clarity. The gray arrows indicate the propagation of the rarefaction solitary wave. } 
\label{fig:Setup}
\end{figure}

 With a firm understanding of the unit cell at hand, we are ready to
 investigate the wave dynamics in a chain of TCO cells (Fig.~\ref{fig:Setup}A).
 The left end of the chain (unit number $n$ = 1) is connected to a shaker through the customized attachment with a sleeve bearing, which transfers
 the shaker impact to the cell while allowing its free rotational motion (see the upper inset of Fig.~\ref{fig:Setup}A).
 The TCO cell positioned in the right end of the chain ($n$ = 20) is fixed to the rigid wall. 
 To obtain the dynamic folding/unfolding motion of each unit cell, we use the digital image correlation (DIC) technique by using six action cameras (GoPro HERO 4 BLACK$^{\tiny{\textregistered}}$) whose maximum frame rate is 240 fps (see the lower inset of Fig.~\ref{fig:Setup}A, and also Materials and Methods for more details). 
 
 The digital images in Fig.~\ref{fig:Setup}B show the snapshots of the first eight TCO unit cells at four different time frames captured by the first pair of the action cameras (see Supplementary Movie S2). In this figure, the length of the colored arrows represents the speed of the polygon in the axial direction, and red (blue) color denotes rightward (leftward) velocity.
 For the sake of visualization, the 3D images are reconstructed based on the experimental data as shown in the right column of Fig.~\ref{fig:Setup}B, where red (blue) color indicates compressive (tensile) strain (see Supplementary Movie S3). Here, the strain value of the $n$th cell is defined by $(u_n-u_{n+1})/h_0$ where $u_n$ ($u_{n+1}$) is the axial displacement of its left (right) polygon, such that compressive strains take positive signs for convenience.  
 From the experimental results, we observe that initially, the first unit shows the large-amplitude compression due to the excitation by the shaker. However, this compressive motion decays quickly without being robustly transmitted along the chain, whereas the noticeable tensile motion is evolved instead (see the arrows in Fig.~\ref{fig:Setup}B, and also Supplementary Movie S2).
 Thus, it appears that a tensile wave has propagated, despite the application of a compressive force. 

\begin{figure} 
\centering
\includegraphics[width=0.9\linewidth]{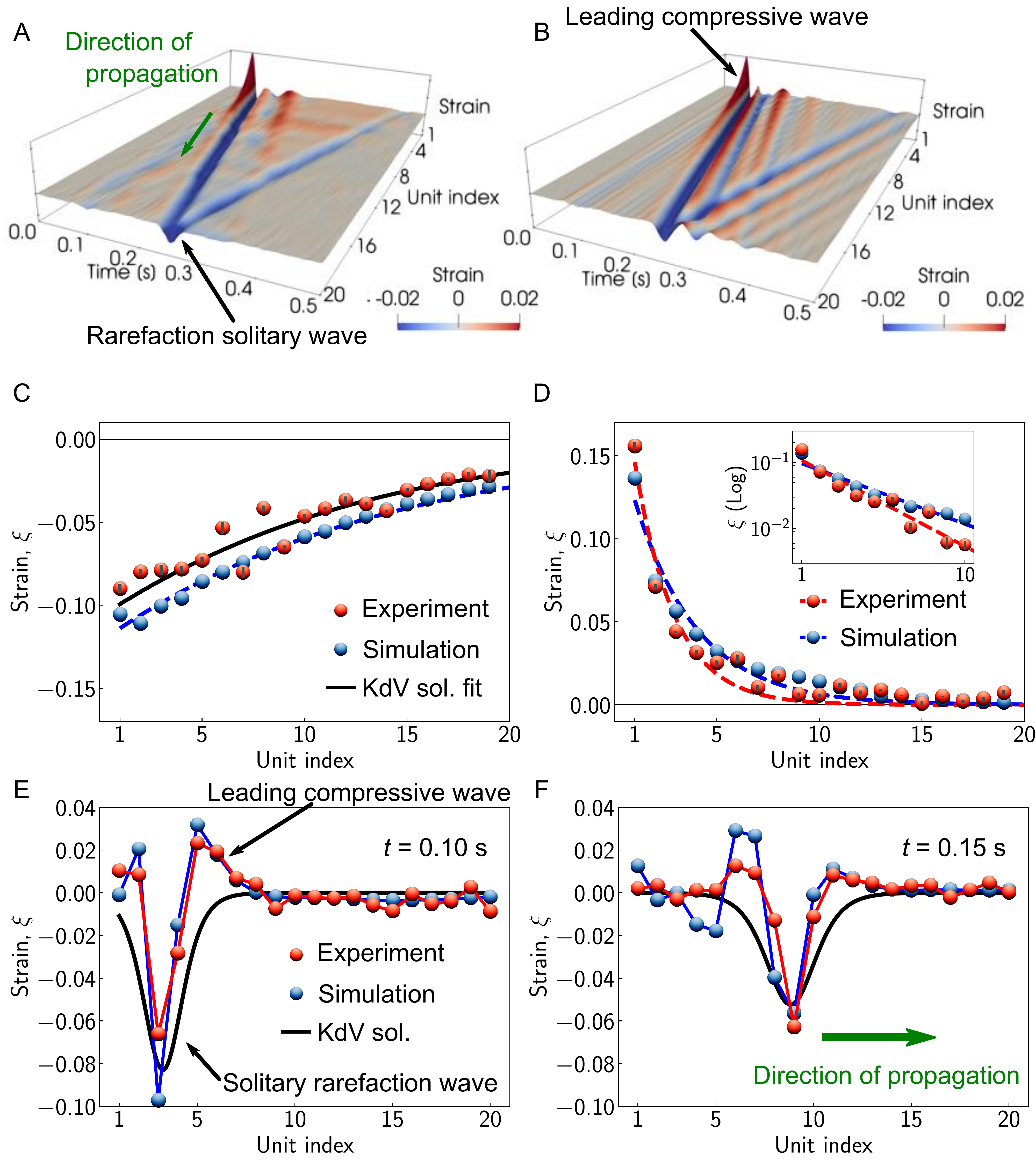}
\caption{ \textbf{Wave form analysis} (\textbf{A}) The space-time
  evolution of the experimentally measured strain wave propagation in the origami-based system. The black arrow indicates the rarefaction solitary wave, and the green shows the direction of the propagation. (\textbf{B}) Numerical simulation results show a qualitative agreement with the experimental data. The black arrow indicates the leading compressive wave in front of the rarefaction wave. (\textbf{C}) The amplitude change of the rarefaction solitary wave. The experimental data is fitted by the KdV solution (black curve) to obtain the damping coefficient for numerical and analytical analysis. The errorbar represents
  the standard deviation (s.d.) calculated from five measurements.  (\textbf{D}) The amplitude of the leading compression is analyzed. The dashed curves are obtained from the exponential fit to the experimental and numerical data. The inset shows the exponential decay of the compressive strain. The shape of the rarefaction solitary wave (\textbf{E}) at $t$ = 0.10 s and (\textbf{F}) $t$ = 0.15 s are shown.} 
\label{fig:Waveform}
\end{figure}

To conduct a more thorough analysis of this counter-intuitive wave dynamics, we plot the measured strains in time and space domains (Fig.~\ref{fig:Waveform}A). We observe that after the compressive impact is applied to the front end of the system, this origami system creates two different types of mechanical waves: (i) small-amplitude and fast-traveling oscillatory waves (see small ripples in $t < 0.2$ s in Fig.~\ref{fig:Waveform}A), and (ii) large-amplitude and slow-traveling, more localized waves (high spikes and dips). The separation of waves into such two groups arises from the two DOF nature of the TCO cells with coupled axial and rotational motions (see the reference~\cite{GilyongLee2016} for wave mixing effects in such settings).  While the former wave group is interesting on its own right, in this study, we focus on the slower but larger-amplitude mechanical waves from a perspective of mitigating impact. In this case, we find interestingly in Fig.~\ref{fig:Setup}A that the primary wave that the last TCO cell ($n=20$) experiences is a tensile wave despite the application of compressive impact to the system.
 
To understand this unique wave dynamics, we now complement the above experimental results with a computational and theoretical analysis. 
 For our computational investigation, we formulate the equations of motion for the entire chain based on the 2DOF model of the TCO cells and solve the resulting equations numerically by using the Runge-Kutta method (see Section II.B in Supplementary Materials for details, and Supplementary Table S1 for the numerical constants used in computational and theoretical analysis). The simulation results are shown in Fig.~\ref{fig:Waveform}B, which demonstrates excellent agreement with the experimental results. 
 
 For our theoretical analysis, we first postulate $u_n \sim \varphi_n$ based on the eigenmode of the single TCO unit cell (see the dashed line in the inset of Fig.~\ref{fig:Static}C). Note that this linear relationship holds approximately in the full range of dynamic strains considered here (see Section II.C and Fig. S5 in Supplementary Materials for an a posteriori validation of this approximation). Based on this, the equations of motion can be reduced to a single component model. Note 
 that the single component model is still nonlinear by virtue of
 the force-displacement and torque-angle relationships.
 We take the continuum limit of the single component equations in the infinite TCO chain in order to derive the well-known KdV equation (see Section II.E in Supplementary Materials for details). This equation has a closed-form
 (rarefaction) solitary wave solution, thereby yielding an analytical approximation
for the wave in the TCO lattice. 

 To account for damping effects in the system (e.g., material dissipation and friction),  we empirically adopt a dashpot damping factor $\nu$
 in the equations of motion. The numerical results in Fig.~\ref{fig:Waveform}B take damping into account. The inclusion of a damping term will modify the derived KdV equation, and the subsequent formula for the rarefaction wave will involve
 the factor $\nu$~\cite{Janaki1992,Ghosh2011}. Using the KdV rarefaction wave as the initial function, we determine $\nu$ = 13.8 s$^{-1}$ by curve-fitting the experimental amplitude of the rarefaction solitary wave in space (red dots in Fig.~\ref{fig:Waveform}C).
  This leads to a good approximation, as indicated by the black solid curve in Fig.~\ref{fig:Waveform}C.
  We find that this results in the reasonably close trend of the strain attenuation in the numerics compared to the experiments as shown in Figs.~\ref{fig:Waveform}, C and D.

 Comparing the attenuation trends of the leading tensile and compressive waves (Figs.~\ref{fig:Waveform}, C and D), we observe the compressive component decays more drastically than the tensile counterpart. The compressive waves show an order-of-magnitude reduction in its amplitude within 10 TCO cells (see the exponential decay of the compressive strain in the inset of Fig.~\ref{fig:Waveform}D).
This manifests the efficacy of the origami-based metamaterial in mitigating the original compressive impact.
  
 In Figs.~\ref{fig:Waveform}, E and F, we show the temporal plots of the strain waves at $t=0.10$ and $0.15$~s. In these plots, the compressive waves are attenuated significantly, while the tensile ones are propagating more dominantly and robustly. Although the analytical waveform focuses only on the tensile components, we observe a qualitative agreement among analytical, numerical, and experimental results.
 One interesting finding here is that the maximum compressive strain is ahead of the rarefaction solitary wave at $t=0.10$ s, but the peak is shifted to the rear part of the solitary wave at $t=0.15$ s. This indicates that the rarefaction solitary wave propagates faster than the other compressive wave packets, and eventually overtakes the initial compressive strain wave. While such overtaking behavior has been previously studied numerically~\cite{Herbold2013,Yasuda2016}, this is the first experimental observation in mechanical platforms, to the best
 of our knowledge.
 

%
%

\begin{figure} 
\centering
\includegraphics[width=1.0\linewidth]{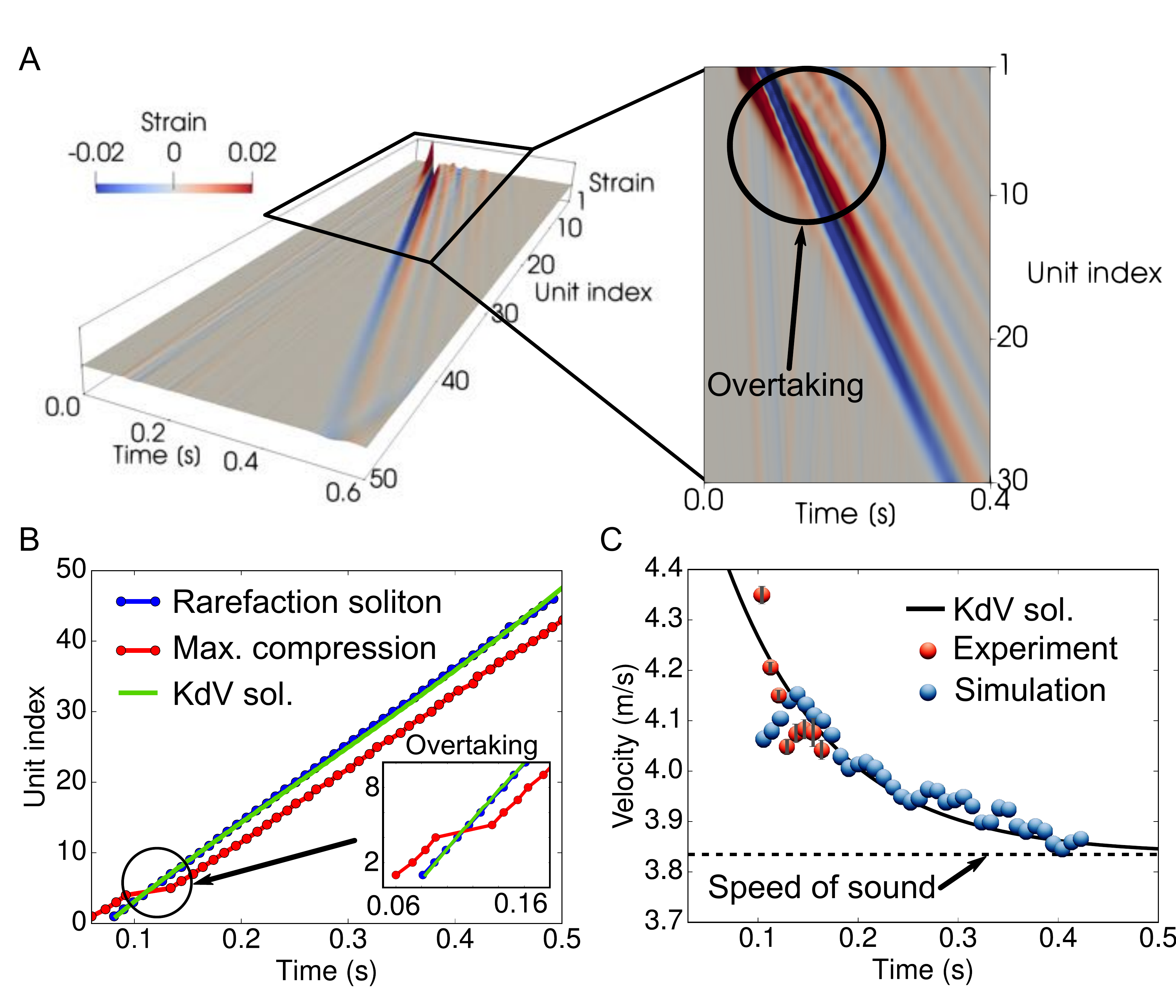}
\caption{ \textbf{Wave speed analysis} (\textbf{A}) Space-time contour plot of the strain wave for the numerical simulation conducted on the longer chain composed of 50 TCO unit cells. Magnified view of the overtaking moment is shown in the right inset. (\textbf{B}) Trajectory of the rarefaction solitary wave (denoted by the blue markers) and the maximum compressive strain wave (red markers) shows the overtaking behavior of the rarefaction solitary wave. The green line indicates the analytical prediction from the KdV equation. (\textbf{C}) Wave speed of the rarefaction solitary wave is larger than the speed of sound of the medium, which means supersonic behavior.}
\label{fig:Velocity}
\end{figure}

 We now further investigate the overtaking of the compressive wave front by the tensile wave (i.e. the rarefaction pulse).  To this end, we conduct numerical simulations for a longer chain composed of 50 unit cells (see Fig.~\ref{fig:Velocity}A and its inset for the magnified view). Upon the compressive impact, the TCO chain initially experiences a compression, followed by the tension resulting from the reaction of the TCO cells. In this case, the speed of the initial (i.e., leading) compressive wave is bounded by
the system's sound speed due to its effectively linear nature. However, the following rarefaction pulse is predicted to be supersonic by the KdV analysis. Thus, we would expect the rarefaction pulse to overtake the compressive linear waves (see Section II.E in Supplementary Materials for details). 
 We observe this overtaking phenomenon around the fifth cell location (see inset of Fig.~\ref{fig:Velocity}A), thereby 
verifying the higher speed of the rarefaction solitary wave than that of the compressive wave pattern. 

To quantify the speeds, we extract the time and location of the maximum peak of the rarefaction solitary and compressive strain waves as shown in Fig.~\ref{fig:Velocity}B.
As marked by the circle in the figure, we detect the overtaking moment around $t=0.11$ s, when the trajectories of solitary wave and maximum compression peak are crossing.
 The slope of each curve represents the wave speed, therefore the wave speed of the rarefaction solitary waves can be calculated from the experimental and numerical results.
 Figure~\ref{fig:Velocity}C shows the wave speed calculated from the experiment (20 units) and simulation (50 units), and the result
 (for the rarefaction solitary wave)
 is compared with the prediction from the KdV solution.
 In addition, the speed of the sound of the system is calculated from the linear analysis (see Section II.E in Supplementary Materials for details). We observe that the experimental and numerical results corroborate the analytical prediction.  In particular, Fig.~\ref{fig:Velocity}C confirms the supersonic nature of the rarefaction pulse.

\section*{Conclusion}
We have studied experimentally, numerically, and analytically
a remarkable example of nonlinear wave propagation  in mechanical metamaterials made of volumetric origami cells.
We found that the TCO-based metamaterials exhibit the rarefaction solitary wave, which features tensile strains and propagates ahead of the initial compressive strain despite the application of external compressive impact.
The KdV-based theoretical analysis provided us with an excellent
qualitative handle for understanding the relevant dynamics, and the
numerical computations corroborated the experimental observations.
  Also, the initial compressive strain is attenuated significantly, which can be highly beneficial for impact mitigation applications.
  While we focused on the monoatomic TCO unit cells with strain-softening behavior in this study, the origami-based system has great potential for supporting rich wave dynamics by introducing heterogeneous elements (e.g., hardening, multi-stable, and impurity components) in the chain. Also, the findings in this 1D setting can be further extended to multi-dimensions in a modular way.  We believe that this architecture of volumetric origami cells can be used as a versatile building block for a wide range of applications such as impact/shock mitigation, vibration filtering, and energy harvesting.



\bibliography{scibib}

\bibliographystyle{Science}

\section*{Acknowledgments}
We thank Professor Sang-gyeun Ahn and Mr. Flyn O'Brien at the School of Art + Art History + Design at the University of Washington for technical support. 
This material is based upon work supported by the National Science Foundation under Grant No. 
CAREER-1553202 (awarded to J.Y.) and DMS-1615037 (awarded to CC). J.Y. and H.Y. are grateful for the support from the ONR (N000141410388).

\section*{Supplementary materials}
Materials and Methods\\
Supplementary Text (sections A to F)\\
Fig.S1. Fabrication of the TCO unit cell.\\
Fig.S2. Compression test on the TCO unit cell.\\
Fig.S3. Digital image correlation (DIC) technique to measure the axial displacement and the rotational angle of each polygon.\\
Fig.S4. Fatigue property of the TCO single unit cell.\\
Fig.S5. Validation of approximation $\varphi_n \sim u_n$.\\
Fig.S6. Comparison between the numerical simulations on the TCO chain with/without damping effect.\\
Fig.S7. Comparison between the numerical simulations on the nonlinear/linearized TCO chain.\\
Table S1. Numerical constants used in the numerical simulation and analytics.\\
Movie S1. Fabrication of the TCO unit cell.\\
Movie S2. Experimental demonstration of the rarefaction solitary wave.
Movie S3. 3D reconstruction of the TCO chain from the experimental result.\\
References(45,46,47,48)


\clearpage



\end{document}